\documentstyle[aps,epsf,psfig,twocolumn]{revtex}
\begin{document}
\draft
\flushbottom
\twocolumn[
\hsize\textwidth\columnwidth\hsize\csname @twocolumnfalse\endcsname

%\draft
\title{ Properties of electrons near a Van Hove singularity}
\author{M. A. H. Vozmediano $^1$, J. Gonz\'alez $^2$,
F. Guinea $^3$, J. V. Alvarez  $^4$, and B. Valenzuela $3$\\}
\address{
         $^1$ Departamento de Matem\'aticas,
         Universidad Carlos III, 
         Avenida de la Universidad 30,
         Legan\'es. 28913 Madrid, Spain\\
	 $^2$Instituto de Estructura de la Materia, 
         Consejo Superior de Investigaciones Cient{\'\i}ficas,
         Serrano 1223, 28006 Madrid, Spain \\
         $^3$Instituto de Ciencia de Materiales, 
         Consejo Superior de Investigaciones Cient{\'\i}ficas, 
         Cantoblanco, 28049 Madrid, Spain \\
	 Universit\H at des Saarlandes,
	 D-66041 Saarbr\H ucken, Germany }
\date{\today}
\maketitle
\widetext
\begin{abstract}

The Fermi surface of most hole-doped cuprates is close to a Van Hove
singularity at the M point.  
A two-dimensional electronic system, whose Fermi surface is close to a
Van Hove singularity shows a variety of weak coupling instabilities. It is
a convenient model to study the interplay between antiferromagnetism and
anisotropic superconductivity. The renormalization group approach is reviewed
with emphasis on the underlying physical processes. General properties of the
phase diagram and possible deformations of the Fermi surface due to
the Van Hove proximity are described.

\end{abstract}

]
\narrowtext 
\tightenlines%\newpage

The Fermi surface of most hole-doped cuprates is close to a Van Hove
singularity at the M point. The possible relevance of this fact to the
superconducting transition as well as to the anomalous behavior of the
normal state was put forward in the early times of the cuprates 
and gave rise to the so-called Van Hove scenario (VHS) \cite{vhscenario}. 
Early critiques to the scenario  have been confronted
with the observation by angle resolved photoemission (ARPES) of very flat 
bands near the Fermi level \cite{vhexp} and with the consistency of most
data with the VHS. Van Hove singularities are found nowadays in a vaste
number of experimental and theoretical papers under various names as
hot spots or flat bands. 

The field theory renormalization group techniques adapted to 
condensed matter problems in \cite{S94}, have been crucial to elucidate
issues such as the renormalization  of the quasiparticle peak and of the Fermi
surface shape as well as to provide a clear picture of the phase diagram in the 
proximity of large peaks in the density of states.

In this paper we give an updated review of the 
VHS and put forward some new features that can be explained within
this framework.

A two dimensional electronic system with a square lattice has typically
two inequivalent high symmetry points in the Brillouin Zone where
a saddle point in the dispersion relation is likely to occur. 
Near a Van Hove singularity the fermion density of states diverges,
so that even arbitrarily weak interactions can produce large effects. 
When the Fermi level reaches these points, a variety of response
functions diverge.

The simplest microscopic model to study this problem is the $t-t'$-Hubbard
model which has the dispersion relation
\begin{eqnarray}
\varepsilon({\bf k})& =& -2t\;[ \cos(k_x a)+\cos(k_y a)] \nonumber \\
&-&2t'\cos(k_x a)\cos(k_y a)-\mu-2t'
\;\;\;, 
\label{disp}
\end{eqnarray}
Band structure calculations \cite{bands}   provide values for the parameters
that can accurately describe the observed Fermi surface of most cuprates. 
For the hole-doped materials, $t'<0$ and, in all cases,
$t'<t/2$. Typical values for BISCO are $t\sim 0.5$ eV, $t'/t \sim -0.3$,
and a small value of t'' that we shall ignore here. 

This dispersion relation has two inequivalent saddle points at 
M $(\pi,0)$ and M' $(0,\pi)$.
The Van Hove model in its simpler 
formulation is obtained by 
assuming that for fillings such that
the Fermi line lies close to the singularities, the majority of
states participating in the interactions will come from regions in the
vicinity of the saddle points. Taylor expanding eq. (\ref{disp}) around
the two points gives the effective relation
\begin{equation}
\varepsilon_{A,B} ( {\bf k} ) \approx \mp ( t \mp 2 t' ) k_x^2 a^2
\pm ( t \pm 2 t' ) k_y^2 a^2 \;\;\;,
\label{vhd}
\end{equation}
which provides a well defined continuous field theory model to which
renormalization group techniques can be rigorously applied \cite{NPB}.

Particle-particle and particle-hole susceptibilities have logarithmic 
divergences giving rise to various instabilities at low
temperatures. The renormalization group allows to study the flow of the
various response functions and to get a phase diagram as a function
of t' and U.

In the simplest of local repulsive interactions,
the leading instabilities are towards ferro- and antiferromagnetism,
and also towards d-wave superconductivity\cite{us98}. 
The phase diagram is shown in fig. 1.

It is interesting to
remark that the electron-hole susceptibility at ${\bf \vec{Q} }
= ( \pi , \pi )$ also diverges logarithmically,
$\chi_{eh} ( {\bf \vec{Q} } ) \sim \log ( \Lambda / \epsilon_F )$,
where $\Lambda$ is a high energy cutoff, of the order of the
bandwidth, $\Lambda \sim 1$eV, and $\epsilon_F$ is the distance 
of the Fermi level to the saddle point, which is typically
one order of magnitude or more smaller.
The ferromagnetic transition, generic of 
two dimensional models of this type\cite{HSG97},
is due to scattering processes which involve small momentum
transfer. These processes are, most likely, screened in
a 3D structure, where scattering with $| {\bf \vec{q}} | \sim
d^{-1}$, where $d$ is the interlayer distance, is 
strongly modified.

From the critical frequencies separating the antiferromagnetic and
the superconducting region in fig. 1, we can see that the critical 
temperature for superconductivity
increases with the value of t', a result in agreement with
recent band theory calculations \cite{andersen}.
\begin{figure}
\centerline{\psfig{file=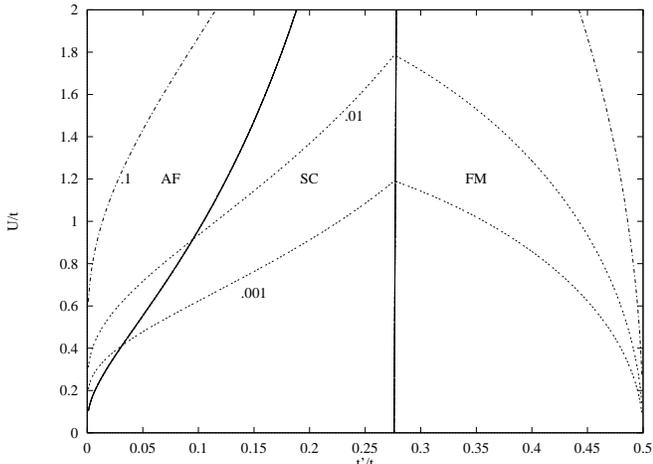,width=3.5in,angle=-90}}
\caption{Phase diagram of the t-t'-Hubbard model near a Van 
Hove singularity from ref. 6.
The dotted lines are contour lines corresponding to the critical frequencies
shown in the figure.}
\end{figure}

Within the same model we can obtain a number of  other features
consistent with the observed behavior of the cuprates:
i) The normal state has anomalous features, including a quasiparticle
lifetime which goes as $| \epsilon |$\cite{LR87,GGV96},
ii) For some values of the parameters, stripes are 
found\cite{VVG00,NK01}, and iii) the out of plane electron hopping
is incoherent\cite{us02}.

The proximity of the Fermi surface to  Van Hove singularities induce
strong renormalization of the Fermi surface shape which can give rise 
to deformations of the Fermi surface breaking the lattice symmetry 
\cite{metzner}. An extended interaction is needed for this to occur 
\cite{belen}.

The competition between d-wave superconductivity
and an instability at ${\bf \vec{Q}}$ leads to three 
possible scenarios:

i) The d-wave susceptibility, which diverges as $\chi_{e-e} (
\omega )
\sim \log ( \Lambda / \omega ) \log ( \Lambda / \epsilon_F )$,
diverges first, before instabilities at ${\bf \vec{Q}}$ develop.
There is a transition to a d-wave superconductor. This
is probably realized in the overdoped regime.

ii) The SDW instability at ${\bf \vec{Q}}$ diverges much faster
than the superconducting one. The ground state
is antiferromagnetic. This is likely to occur
near half filling.

iii) Both instabilities are comparable. The ground state will be,
most likely, a  d-wave superconductor. The reason is that, in
the absence of perfect nesting, the instability at ${\bf \vec{Q}}$,
even if fully developed,
does not lead to an insulating state. Portions of the Fermi
surface remain gapless. The logarithmic
divergences in the BCS channel are not fully suppressed,
and become $\chi_{e-e} (
\omega )
\sim \log ( \Lambda / \omega ) \log ( \Lambda / 
\Delta_{\bf \vec{Q}} )$, where $\Delta_{\bf \vec{Q}}$ is a
scale determined by the divergence of $\chi_{e-h}
( {\bf \vec{Q}} )$. The existence of anomalous behavior
at a scale above the superconducting critical temperature
has been widely reported in the underdoped regime,
and is usually associated to the pseudogap.

In the case where the Fermi surface does not have space symmetries,
the phase diagram shows competition between ferromagnetism and p-wave
superconductivity \cite{us00}. Coexistence of ferromagnetism and triplet
superconductivity has been reported in $MgCNi_3$ \cite{mg}
whose Fermi surface lies close to Van Hove singularities \cite{scMg},
and in the ruthenium compounds \cite{rute}.

As a summary we list the main results obtained by the combined RG techniques 
in the VHS which have some  experimental support.

- vanishing of the quasiparticle peak ,

- pinning of the Fermi surface at the singularity,

- in-plane resistivity linear in frequency,

- incoherence of the c axis transport, 

- deformation of the Fermi surface breaking the tetragonal symmetry of the lattice
(Pomeranchuk instability),

- coexistence of antiferromagnetism d-wave superconductivity, and 
existence of a ferromagnetic phase if the Fermi surface 
is four fold symmetric,

- increasing of $T_c$ with $t'$ in the t-t' model,

- coexistence of ferromagnetism and p wave superconductivity for asymmetric
Fermi surfaces.


\begin{references}

\bibitem{vhscenario} 
J. Labb\'e and J. Bok, {\em Europhys. Lett.} {\bf 3} (1987) 1225.
J. Friedel, {\em J. Phys.} (Paris) {\bf 48} (1987) 1787; {\bf 49}
(1988) 1435.
H. J. Schulz, {\em Europhys. Lett.} {\bf 4}, 609 (1987).
R. S. Markiewicz and B. G. Giessen, {\em Physica} (Amsterdam)
{\bf 160C} (1989) 497.
R. S. Markiewicz, {\em J. Phys. Condens. Matter} {\bf 2} (1990) 665.
D. M. Newns {\em et al.}, {\em Phys. Rev. Lett.} {\bf 69}, 1264 (1992).

\bibitem{vhexp} D. S. Dessau {\it et al.}, {\em Phys. Rev. Lett.} 
{\bf 71}, 2781 (1993), D. M. King  {\it et al.}, {\em Phys. Rev. Lett.} 
{\bf 73}, 3298 (1994),  K. Gofron {\it et al.}, {\em Phys. Rev. Lett.} 
{\bf 73}, 3302 (1994). See Z.-H. Shen  {\it et al.}, Science {\bf 267}, 343
(1995), and references therein.

\bibitem{S94}
R. Shankar, {\em Rev. Mod. Phys.} {\bf 61}, 433 (1989). 

\bibitem{bands} W. E. Pickett,  {\em Rev. Mod. Phys.} {\bf 66}, 129 (1994).

\bibitem{NPB} J. Gonz\'alez, F. Guinea and M. A. H. Vozmediano,
{\em Nucl. Phys.} {\bf B485}, 694 (1997).

\bibitem{us98} J. V. Alvarez, J. Gonz\'alez, F. Guinea and M. A. H. Vozmediano,
{\em J. Phys. Soc. Jpn.} {\bf 67}, 1868 (1998).

\bibitem{HSG97}
R. Hlubina, S. Sorella and F. Guinea, Phys. Rev. Lett. {\bf 78}, 1343
(1997).

\bibitem{andersen} E. Pavarini, I. Dagupta, T. Saha-Dagupta, O. Jepsen,
and O. K. Andersen, this proceedings.

\bibitem{LR87}
P. A. Lee and N. Read, Phys. Rev. Lett. {\bf 58}, 2691 (1987).

\bibitem{GGV96}
J. Gonz\'alez, F. Guinea and M. A. H. Vozmediano, Europhys. Lett.,
{\bf 34}, 711 (1996).

\bibitem{VVG00}
B. Valenzuela, M. A. H. Vozmediano, and F. Guinea, Phys. Rev. B
{\bf 62}, 11312 (2000).

\bibitem{NK01}
B. Normand and A. Kampf, preprint (cond-mat/0102201).

\bibitem{us02}
M. A. H. Vozmediano, F. Guinea and M. P. L\'opez-Sancho,
to be published.

\bibitem{us00} J. Gonz\'alez, F. Guinea and M. A. H. Vozmediano,
{\em Phys. Rev. Lett.} {\bf 84}, 4930 (2000).

%\bibitem{pfeuty} F. Onufrieva and P. Pfeuty, 
%{\em Phys. Rev. B.} {\bf 61}, 799 (2000).

\bibitem{HO00}
A. Himeda and M. Ogata, preprint (cond-mat/0003278).


\bibitem{metzner} C. J. Halboth and W. Metzner,  
{\em Phys. Rev. Lett.} {\bf 85}, 5162 (2000).

\bibitem{belen}  B. Valenzuela and M. A. H. Vozmediano, 
{\em Phys. Rev. B.} {\bf 63}, 153103  (2001).


\bibitem{mg} T. He,  {\it et al.}, {\em Nature}  {\bf 411}, 54 (2001).

\bibitem{scMg} H. Rosner,  {\it et al.}, preprint (cond-mat/0106583).

\bibitem{rute} R. Matzdorf, {\it et al.}, {\em Science} {\bf 289}, 746 (2000);
K. M. Shen,  {\it et al.}, preprint (cond-mat/0105487).

\end{references}
\end{document}